\documentclass[fleqn,usenatbib]{mnras}

\usepackage{newtxtext,newtxmath}

\usepackage[T1]{fontenc}

\DeclareRobustCommand{\VAN}[3]{#2}
\let\VANthebibliography\thebibliography
\def\thebibliography{\DeclareRobustCommand{\VAN}[3]{##3}\VANthebibliography}

\usepackage{graphicx}	
\usepackage{amsmath}	

\title[Role of EoS on spin-up of MSPs]{The role of equation of state on the spin-up of millisecond pulsars}

\author[X. Y. Zhong et al.]{
Xinyi Zhong,$^{1}$
Xiaoyu Lai,$^{2}$\thanks{E-mail: laixy@hue.edu.cn (XYL)}
\\
$^{1}$School of Physics and Optoelectronic Engineering, Yangtze University, Jingzhou 434023, China\\
$^{2}$Institute of Astronomy and High Energy Physics, Hubei University of Education, Wuhan 430205, China
}

\date{Accepted XXX. Received YYY; in original form ZZZ}

\pubyear{2015}

\begin{document}
\label{firstpage}
\pagerange{\pageref{firstpage}--\pageref{lastpage}}
\maketitle

\begin{abstract}
Millisecond pulsars (MSPs) are recycled pulsars which have been spun-up due to mass accretion during the phase of mass exchange in binaries.
Although the interactions with companion stars play important roles on the spin-up process, the global properties of pulsars determined by the equation of state (EoS), such as mass, radius and the moment of inertia, should also play a role.
We investigate the spin-up of MSPs in neutron star (NS) and strangeon star (SS) models, both of which have passed the tests by the existence of high-mass pulsars and the tidal deformability of GW~170817.
Combining the spin-up condition and the transferred angular momentum, and taking into account the evolution of magnetic field strength during accretion, we can constrain the spin-period and mass of an MSP.
Our results show that the impeding effect of magnetic field on the spin-up of MSPs would be more significant for NSs than for SSs, especially for the ones with low masses.
In the low-mass ($M$ below or around about $1.2 M_\odot$) case, an SS can spin faster than an NS of the same mass by accreting the same amount of mass.
Finding more low-mass and fully recycled MSPs, with accurate mass-measurement and better constraints on the amount of accreted mass, could help to put more strict constraints on the EoS of pulsars.
\end{abstract}




\section{Introduction} \label{sec:intro}

The pulsar-like compact stars are idea laboratories to explore the state of matter at supra-nuclear densities.
The equation of state (EoS) of pulsars is still a controversial topic at present, although abundant observational data of pulsars has been accumulated.
Based on different points of view, a variety models for pulsars have been speculated.
The neutron star (NS) model whose basic idea could be found in~\citet{Landau1932} and the quark star (QS) model which originated from Witten's conjecture~\citep{Bodmer1971,Witten1984} have attracted most attention.
The strangeon star (SS) model was also proposed~\citep{Xu2003} to understand different manifestations of pulsars, where the strong coupling between quarks has grouped quarks into strangeons (i.e. ``strange nucleons'') , and has passed various observational tests~\citep[see][and references therein]{LXX2023} 

The EoS of pulsars is essentially depends on the behavior of quantum chromodynamics (QCD) at low-energy scales, which is still a challenge for us to understand.
Although the constraints by both the existence of high-mass pulsars and the tidal deformability of GW 170817~\citep{GW170817} could dramatically reduce the allowed EoSs of NSs~\citep{Annala2017}, it is still uncertain that whether pulsars are NSs, QSs or SSs.
In addition, these three kinds of models mentioned above have subclass models respectively, e.g., some branches of NS model still allow the existence of quark matter at the core region~\citep{Weber2005}, QS model includes different states such us unpaired strange quark matter state~\citep{Alcock1986} and colour super-conductivity (CSC) state~\citep{Alford2008}, and the hybrid SS model is proposed~\citep{Zhang2023} where an SS develops a strange quark matter core.
Therefore, the differences among models could be subtle.
It is certainly important to find out unambiguous and observable probes to tell what is the real nature of pulsars.

The global structure of pulsars, e.g. the mass and radius, can provide robust constraints on EoS models when compared with observations~\citep{Lattimer2007,Ozel2010}.
Despite the uncertainty about the explicit structure determined by various classes of EoS models, there are some general differences in the global structures between NSs and QSs/SSs~\citep{Demorest2010,Gao2022}.
NSs are gravity-bound so that the surface density could approach zero, and the radii of low-mass NSs (below $\sim M_\odot$) decrease with increasing masses and the radii of intermediate-mass NSs (in the range between $\sim M_\odot$ and $\sim 2M_\odot$) could remain nearly constant.
Both QSs and SSs are self-bound by strong interaction so that the surface density could still above the nuclear matter density, and their radii increase with increasing masses unless the mass approaches its maximum value.
The distinctions due to gravity-binding and self-binding, especially for the mass below $\sim 1.4 M_\odot$ when QSs/SSs would be more compact than NSs with the same mass, seem to be more pronounced than that due to, e.g., whether the degrees of freedom are free quarks or strangeons, or whether quark matter emerges inside NSs or not.
Therefore, testing whether pulsars are gravity-bound or self-bound provides a relatively less unambiguous probe to the nature of pulsars.

In this paper we investigate that how different global structures of pulsars from different EoS models affect the spin-up process of millisecond pulsars (MSPs).
MSPs evolve from pulsars in binaries which have underwent the spin-up process, i.e. recycling process, due to accreting mass from their companions.
The formation of MSPs with a carbon-oxygen white dwarf (CO WD)~\citep{Tauris2000,Tauris2011} or a helium white dwarf (He WD)~\citep{Tauris2012,Chen2021} in intermediate-mass X-ray binaries (IMXBs) and low-mass X-ray binaries (LMXBs)~\citep{Podsiadlowski2002} have been investigated to show that the spin-period of a recycled MSP is related to the amount of accreted mass which depends on the nature of the donor star.
The observed differences of the spin distributions of recycled pulsars with different types of companions have been explained by the amount of accreted mass needed for a pulsar to achieve its equilibrium spin~\citep{Tauris2012}.
In addition, the magnetic field evolution of accreting pulsars would also affect the formation of MSPs.
The spin-period evolution of recycled pulsars could have been influenced by the accretion induced magnetic field decay~\citep{Zhang1998,Zhang2006}.
The minimum spin-period of millisecond pulsars also depends the correlation between the accretion rate and the surface dipole magnetic field in LMXBs~\citep{Alpar2021}.

The spin-up line in the $P\dot{P}$-diagram is often used to demonstrate the formation and evolution of recycled pulsars, although it in fact cannot be uniquely defined~\citep{Tauris2012} due to the dependence on both the process of accretion and the properties of the pulsar.
Most of the previous studies about the evolution of MSPs focusing on the accretion process in the binary evolution use the typical values for the structure of neutron stars.
These studies discuss many parameters which do affect the location of the spin-up line in the $P\dot{P}$-diagram, e.g. the critical fastness parameter~\citep{Wang2011} and the location of magnetospheric boundary~\citep{Ghosh1979b}.
It is worth noting that, the spin evolution of a pulsar also depends on the EoS which determines the pulsar's global structure such as mass, radius and the moment of inertia.
In this paper we will investigate the role of equation of state on the spin-up of MSPs.

The EoSs should satisfy the constraints by both the existence of higher than two-solar-mass pulsars and the tidal deformability of GW 170817. In this paper we use two kinds of EoS models for strangeon stars (SSs): LX3630~\citep{LX2009b,Gao2022} and Z-2023 (hybrid SSs)~\citep{Zhang2023}. We also use two kinds of EoS models for neutron stars (NSs): AP4~\citep{AP1997} and X-2024 (hybrid NSs)~\citep{Xia2024}. A hybrid SS is an SS which develops a strange quark matter core, so it is self-bound. A hybrid NS is an NS which develops a strange quark matter core, so it is gravity-bound. Therefore, we choose SSs in both LX3630 and Z-2023 models to represent self-bound SSs, and choose NSs in both AP4 and X-2024 to represent gravity-bound NSs.
Considering the large error of NICER measurements (e.g.~\citet{Miller2021}), we do not use their results to constrain EoS models.

\begin{figure}
\includegraphics[width=\columnwidth]{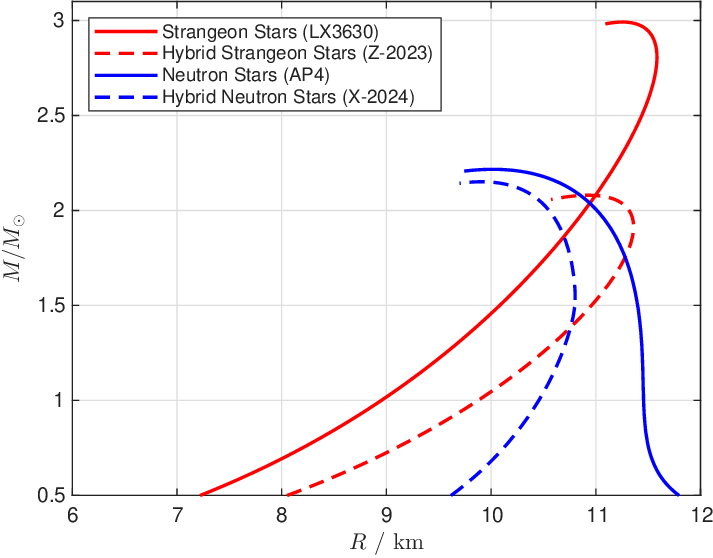}
\caption{$M$-$R$ curves for SSs in LX3630 (red solid lines), hybrid SSs in Z-2023 (red dashed line), NSs in AP4 (blue solid line), and hybrid NSs in X-2024 (blue dashed line). In this paper we use the former two models to describe SSs and the latter two models to describe NSs} \label{fig:MR}
\end{figure}

From the mass ($M$)-radius ($R$) curves shown in Fig.\ref{fig:MR}, we can see that for the mass below $\sim 1.4 M_\odot$ SSs would be more compact than NSs with the same mass. SSs could have very low masses and small radii, and the discrepancy between the results of LX3630 and AP4 are more obvious. We try to demonstrate that, for the spin-up process of MSPs there could be observational differences between SSs and NSs in the low-mass case.

From the spin-up condition we derive the spin-up lines in $MP$ (mass vs spin-period) diagram in Sec.\ref{sec:spin-up}, and find that under the same magnetic field strength an SS could spin faster than an NS in the recycling process.
To investigate the effect of accretion process, in Sec.\ref{sec:accretion} we derive the spin-period as the function of accreted mass, taking into account the evolution of magnetic field strength during accretion.
Then we can constrain the spin-period and mass of an MSP.
The $M$-$P$ lines shown in Sec.\ref{sec:result} indicate that the constraints on EoS models could be better for masses below or around about $1.2 M_\odot$, especially when we could get more accurate constraints on the amount of accreted mass.
Conclusions and discussions are made in Sec.\ref{sec:summary}.

\section{The spin-up line} \label{sec:spin-up}

Consider an accreting compact star with mass $M$, accretion rate $\dot{M}$, radius $R$, spin-period $P$ and surface dipole magnetic field $B$.
The Alfv$\acute{\rm e}$n radius $r_{\rm A}=[B^2R^6/(\dot{M}\sqrt{2GM})]^{2/7}$ is defined as the location where the magnetic energy density equals to the kinetic energy density of the accreted matter.
The magnetospheric radius $r_{\rm m}$ is the location inside which the falling particles will flow along the magnetic field lines.
Generally, $r_{\rm m}=\phi \cdot r_{\rm A}$ due to the influence of the disk-fed accretion flow, where the factor $\phi$ depends on models describing the inner edge of the disc. \citet{Ghosh1979a} found $\phi\approx 0.5$,  \citet{Burderi1998} adopted $\phi=1$, and \citet{Bozzo2009} set $\phi$ to be about 0.5-1.4.

Due to the model-dependence of the accretion torque on the spin-period and the magnetic field configuration~\citep{Ghosh1979b}, to be spun-up by the accretion the pulsar should have the spin angular velocity $\Omega$ satisfying $\Omega\leq\omega_{\rm c}\cdot\Omega_{\rm K}(r_{\rm m})$, where $\Omega_{\rm K}(r_{\rm m})$ is the Keplerian angular velocity at radius $r_{\rm m}$, and $\omega_{\rm c}$ is the critical fastness parameter.
To exert a positive accretion torque $\omega_{\rm c}$ should be about 0.2-1~\citep{Ghosh1979a}, and the spin frequency of 716 Hz could indicate that $\omega_{\rm c}=0.55$~\citep{Wang2011}.

The spin-up condition can therefore be expressed as the relation between $r_{\rm A}$ and $r_{\rm co}$, where the corotation radius $r_{\rm co}=[GMP^2/(4\pi^2)]^{1/3}$ is defined as the radius where the Keplerian frequency matches the spin frequency of the pulsar.
Using the factors $\phi$ and $\omega_{\rm c}$ mentioned before, the spin-up condition can be written as
\begin{equation}
r_{\rm A}\leq [\omega_{\rm c}^{2/3}/\phi] r_{\rm co}. \label{eq:spin-up}
\end{equation}
Because we will focus on the EoS-dependent results of the spin-up process, the dependence on other factors, such as truncation of the accretion disc and the magnetic field configuration outside the star, will not be included when we are comparing different EoS models.
In the following we adopt $\phi=0.7$ and $\omega_{\rm c}=0.55$.

The strength of the magnetic dipole field $B$ is usually estimated in terms of the spin period $P$ and its time derivative $\dot{P}$, by applying the vacuum magnetic dipole model and assuming that the rate of rotational energy loss equals the energy-loss rate caused by dipole radiation due to the inclination angle $\alpha$ (between the axis of the magnetic dipole field and the rotation axis of the pulsar).
However, the real spin-down torque should be more complicated than the magnetic dipole braking in vacuum.
For example, the spin-down torque could be exerted by particle wind~\citep{Li2014,Tong2016} or by the plasma current in the magnetosphere~\citep{Spitkovsky2006}.
More importantly, the value of $B$ calculated from $P$ and $\dot{P}$ under any spin-down torque model should be very different from that in the accreting process which we will consider in Sec.\ref{sec:accretion}.
Therefore, we keep $B$ as a parameter and plot the spin-up line in $MP$ diagram.
Then assuming the mass accretion rate $\dot{M}=0.1\dot{M}_{\rm Edd}$ (the Eddington accretion limit $\dot{M}_{\rm Edd}\simeq 2\times 10^{-8} M_\odot \rm yr^{-1}$), and using equation~(\ref{eq:spin-up}), we can plot the spin-up line in $MP$ diagram.

\begin{figure}
\includegraphics[width=\columnwidth]{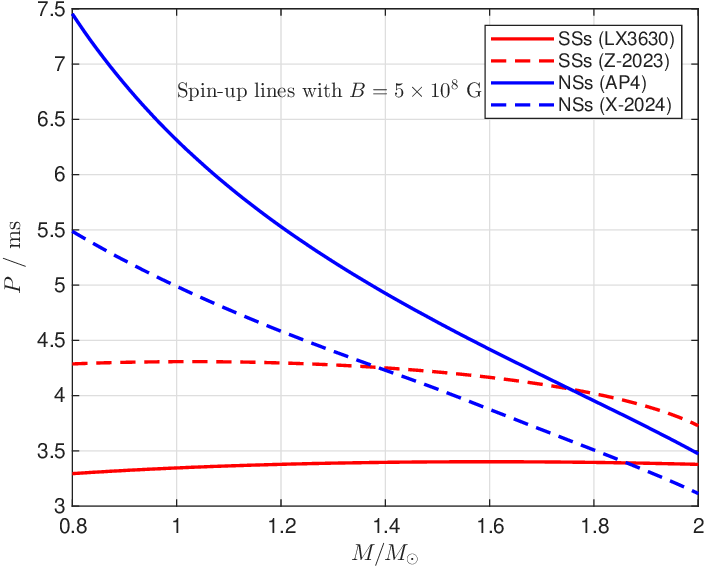}
\caption{Spin-up lines of $B=5\times 10^8$ G for SSs in LX3630 (red solid line) and Z-2023 (red dashed line), and for NSs in AP4 (blue solid line) and X-2024 (blue dashed line).} \label{fig:spin-up}
\end{figure}

The spin-up condition written in equation~(\ref{eq:spin-up}) is an equation connecting mass and radius, so it is EoS-dependent.
Different EoS models will give different spin-up lines in $MP$ diagram, and Fig.\ref{fig:spin-up} shows the results for SSs (including LX3630 and Z-2023 models) and NSs (including AP4 and X-2024 models).
The four spin-up lines of $B=5\times 10^8$ G correspond to SSs in LX3630 (red solid line), SSs in Z-2023 (red dashed line), NSs in AP4 (blue solid line), and NSs in X-2024 (blue dashed line), respectively.
We can see that the spin-up lines of NSs (both AP4 and X-2024 models) are decreasing with the mass, and that of SSs (both LX3630 and Z-2023 models) are nearly independent of the mass.

The spin-up line gives the lower limit for spin-period $P$ of recycled pulsars, so Fig.\ref{fig:spin-up} indicates that by a sufficient mass-transfer an SS could have smaller values of $P$ than an NS in the low-mass case.
The dashed lines in Fig.~\ref{fig:spin-up} show that the magnetic field strength will play an important role in impeding the spin-up of NSs, especially for the ones with masses. When below about $M< 1.4 M_\odot$, the impeding effect would be less important for SSs than for NSs with the same values of $B$.
This is understandable: with the same mass below $1.4 M_\odot$, the radius of an NS is usually larger than that of an SS, i.e. an NS usually has larger $r_{\rm A}$ than an SS, which means that in the low-mass case it is more difficult to satisfy the spin-up condition (\ref{eq:spin-up}) for an NS than that for an SS.

The difference between SSs in LX3630 (red solid line) and NSs in AP4 (blue solid line) is more pronounced than that between SSs in Z-2023 (red dashed line) and NSs in X-2024 (blue dashed line).
This can be inferred from Fig.\ref{fig:MR}, which shows that at low masses, the discrepancy of $M$-$R$ curves between LX3630 and AP4 models is more obvious than that between Z-2023 and X-2024 models.

\section{Spin-periods after accretion} \label{sec:accretion}

By accreting the amount of mass ${\rm d} M$, the spin angular momentum added to a pulsar is given by
\begin{equation}
{\rm d} J=n \sqrt{GMr_{\rm m}}{\rm d} M, \label{eq:P}
\end{equation}
where $n$ is a dimensionless torque.
In principle, we can derive the final spin-period of a pulsar (from rest) after accreting the amount of mass $\Delta M = M-M_0$ (if the spin-up condition is satisfied).

When considering the accretion process, two factors should be taken into account. (1) The torque $n$ depends on fastness parameter ($\Omega/\Omega_{\rm K}(r_{\rm m})$).
We use the formula $n=1-\Omega/\Omega_{\rm K}(r_{\rm m})$ from~\citet{Menou1999}.
This indicate that the torque decreases as the spin-up proceeds.
(2) The evolution of magnetic field $B$ during accretion.
The accretion leads to decay of magnetic fields, which then changes the magnetospheric radius $r_{\rm m}$.
We use the relation between $B$ and the accreted mass $\delta M$ given by~\citet{Zhang2006}, $B\simeq 0.8 B_{\rm f}(\delta M/M_{\rm cr})^{-1.75}$, where $B_{\rm f}=10^8$ G and $M_{\rm cr}=0.2M_\odot$.

The spin-period with the amount of accreted mass $\Delta M$ derived from equation~(\ref{eq:P}) is also EoS-dependent.
The angular momentum $J=I\Omega$, where $I$ is the moment of inertia.
The values of $I$ are derived when the spin angular momentum of the star is calculated to the first order of spin angular velocity $\Omega$~\citep{Gao2022}, which means that $I$ is independent of $\Omega$, and this approximation is good when the $P>1$ ms.
Then by assuming $\phi=0.7$ and $\dot{M}=0.1\dot{M}_{\rm Edd}$ as in Sec.\ref{sec:spin-up}, equation~(\ref{eq:P}) is related to $M$, $R$ and $I$.
Given an EoS, the spin-period $P$ with the amount of accreted mass $\Delta M$ could be derived.

\begin{figure}
\includegraphics[width=\columnwidth]{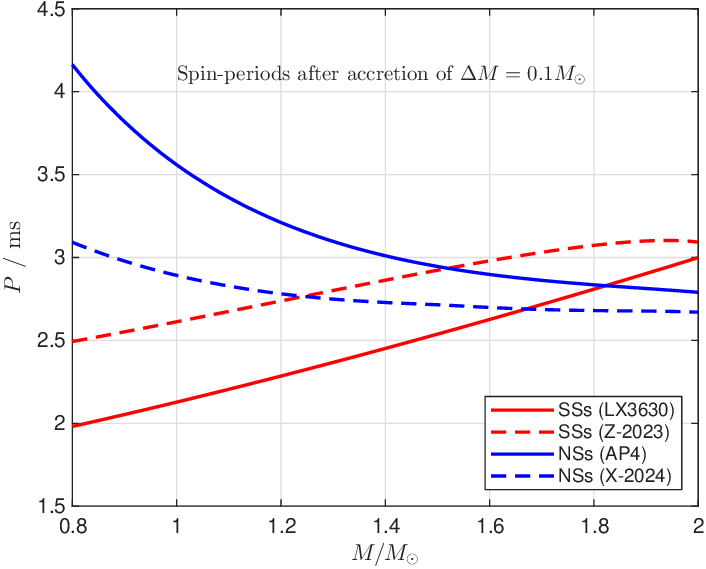}
\caption{The curves of $P$ after accretion of the amount of mass $\Delta M=0.1 M_\odot$, as the function of the final mass $M$. The results of SSs in LX3630 and Z-2023 models are shown respectively by red solid line and red dashed line, and the results of NSs in AP4 and X-2024 models are shown respectively by blue solid line and blue dashed line. } \label{fig:accretion}
\end{figure}

The curves of spin-period $P$ after accretion of the amount of mass $\Delta M=0.1 M_\odot$ are shown in Fig.\ref{fig:accretion}, where the horizontal ordinate denotes the final mass after accretion.
The results of SSs in LX3630 and Z-2023 models are represented respectively by red solid line and red dashed line, and the results of NSs in AP4 and X-2024 models are represented respectively by blue solid line and blue dashed line.
We can see that, with the same amount of accreted mass, in the low-mass case SSs would spin faster than NSs.
Larger/smaller values of $\Delta M$ will move the lines downwards/upwards.

The fastness-dependent torque $n$ indicates that the spin-up itself will impede the further spin-up.
The difference in tendencies of $P$ towards increasing mass between NSs and SSs shown in Fig.\ref{fig:accretion} reflects the difference in such kind of impediment for NSs and SSs.
When $M<1.4 M_\odot$, an NS usually has a larger $r_{\rm m}$ than that of an SS with the same mass, which means that in the low-mass case the impediment of spin-up is more significant for NSs than that for SSs.
This difference essentially reflects the distinction between stellar structures of gravity-bound NSs and self-bound SSs.

\section{Comparison with observations} \label{sec:result}

The recycling of pulsars is usual studied by comparison with observations in $P\dot{P}$ diagram.
As we mentioned above, we keep $B$ as a parameter instead of replacing it by combination of $P$ and $\dot{P}$ to derive the spin-up lines in $MP$ diagram, and take into account the evolution of $B$ in the spin-up process.
Therefore, as shown in equations (\ref{eq:spin-up}) and (\ref{eq:P}), in this paper the quantities to be compared with observations are $M$ and $P$.

The spin-up lines derived in Section \ref{sec:spin-up} do not consider the supplied mass, and the spin-periods after accretion of $\Delta M$ derived in Section \ref{sec:accretion} do not consider the spin-up condition, so we can derive the spin-periods after accretion of $\Delta M$ by putting Fig.\ref{fig:spin-up} and Fig.\ref{fig:accretion} together.
In addition, because we consider the evolution of $B$ during accretion, the spin-up lines depending on the value of $B$ can be determined by the accreted mass.

\begin{figure}
\includegraphics[width=\columnwidth]{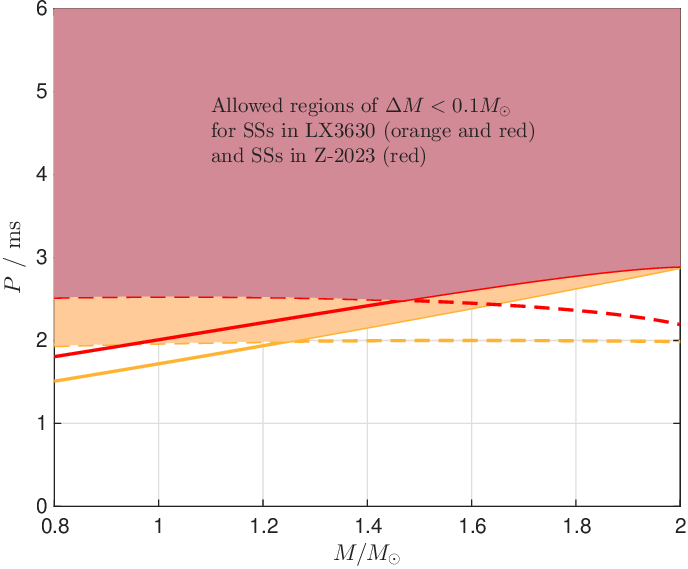}
\includegraphics[width=\columnwidth]{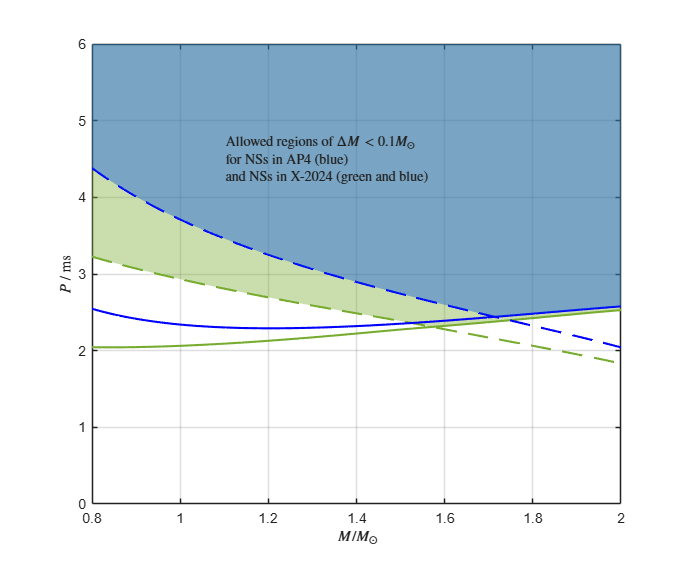}
\caption{The allowed regions of SSs (upper) and NSs (lower) in $MP$ diagram. The upper panel shows the spin-periods of SSs with mass $M$ after accretion of $\Delta M=0.1 M_\odot$ in LX3630 (orange solid line) and Z-2023 (red solid line), and the spin-up line of SSs whose magnetic field strength equals to the final $B$ after accretion of $\Delta M=0.1 M_\odot$ in LX3630 (orange dashed line) and Z-2023 (red dashed line). If $\Delta M<0.1 M_\odot$, the allowed region for SSs in LX3630 is covered by both orange and red patches, and that for SSs in Z-2023 is covered by red patch. The lower panel of shows the same results as in the upper one, for NSs in AP4 model (blue solid line, blue dashed line, and blue patch) and in X-2024 model (green solid line, green dashed line, and patches in both blue and green).} \label{fig:allowed}
\end{figure}

In the upper panel of Fig.\ref{fig:allowed}, the orange solid line shows the spin-periods of SSs in LX3630 model with mass $M$ after accretion of $\Delta M=0.1 M_\odot$, and the orange dashed line is the spin-up line of SSs in LX3630 whose magnetic field strength equals to the final $B$ after accretion of $\Delta M=0.1 M_\odot$.
Therefore, when $\Delta M=0.1 M_\odot$, the spin-period of an SS in LX3630 lies in the curve formed by taking the higher parts of these two lines.
The red solid and dashed lines are the corresponding results of SSs in Z-2023 model.
From the observational point of view, it would be easier to constrain the upper limit of $\Delta M$, then we can get the allowed region which lies above both solid and dashed lines.
We can see that the allowed region for SSs (covered by both orange and red patches) in LX3630 is bigger than and encloses that for Z-2023 (covered by red patch).

The lower panel of Fig.\ref{fig:allowed} shows the same lines as in the upper one, for the results of NSs in AP4 model (blue solid and dashed lines) and X-2024 model (green solid and dashed lines).
The allowed region for X-2024 model (covered by both green and blue patches) is bigger than and encloses that for AP4 model (covered by blue patch).
Comparing the upper and lower panels of Fig.\ref{fig:allowed} we can see that, the allowed region of SSs (in both LX3630 and Z-2023 models) covers the lower left region in $MP$ diagram where NSs (in both AP4 and X-2024 models) cannot reach, indicating that it could be easier for an SS to be spun-up to a period less than 3 ms than that for an NS, when the mass is below about $1.2 M_\odot$.

As $\Delta M$ increases/decreases, all the lines for $P$ after accretion of $\Delta M$ (solid lines) and spin-up lines (dashed lines) in Fig.\ref{fig:allowed} will shift downwards/upwards.
The allowed region of each model will be above both the solid and dashed lines in the corresponding color.
For some values of $\Delta M$, solid line will be totally above or below the dashed line.
In all cases, for a certain value of $\Delta M$ the spin-period of an MSP lies in the curve formed by taking the higher parts of solid and dashed lines in the same color, and for an upper limit of $\Delta M$ the allowed region of an MSP lies above both solid and dashed lines in the same color.

The above example shows that, the role of EoS on the spin-up of MSPs could be manifested in low-mass (below about $1.2 M_{\odot}$) case.
As also shown in Fig.\ref{fig:spin-up} and Fig.\ref{fig:accretion}, the difference between SSs in LX3630 and NSs in AP4 is more pronounced than that between SSs in Z-2023 and NSs in X-2024 model.

\begin{figure}
\includegraphics[width=\columnwidth]{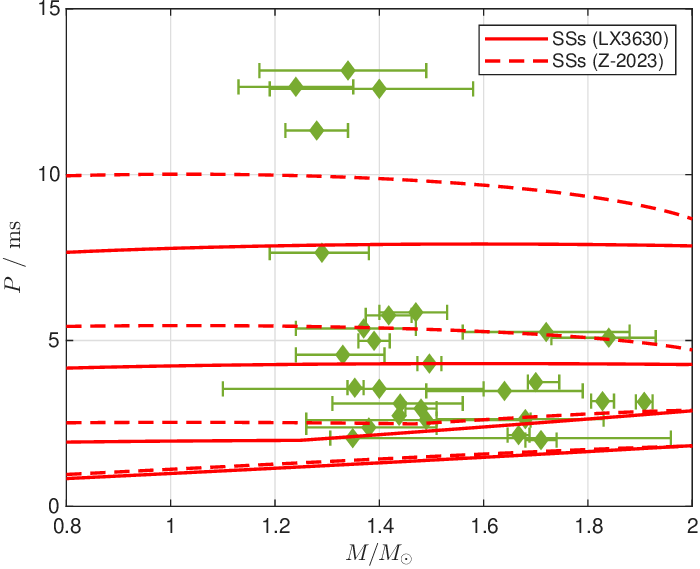}
\includegraphics[width=\columnwidth]{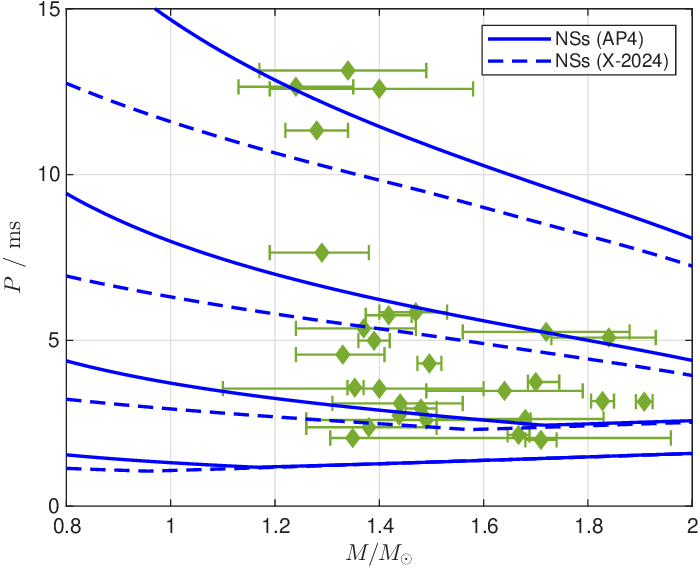}
\caption{$M$-$P$ curves after accretion of $\Delta M$, for SSs in LX3630 (red solid line in upper panel), SSs in Z-2023 (red dashed line in the upper panel), NSs in AP4 (blue solid line in lower panel), and NSs in X-2024 (blue dashed line in lower panel). For each model, we show the results for $\Delta M/M_\odot=0.04, 0.06, 0.1, 0.2$, increasing downwards. Data points with 1-$\sigma$ uncertainties are from~\citet{Ozel2016}.} \label{fig:MP}
\end{figure}

To compare with observational data, we extend the curves in Fig.\ref{fig:allowed} to different values of $\Delta M$, and we show in Fig.\ref{fig:MP} the results for SSs (upper) and NSs (lower).
Each line is derived by taking the higher parts of line of $P$ after accretion of $\Delta M$ and line of spin-up, for a certain value of $\Delta M$.
In the upper panel, the results for SSs in SS model (LX3630) and hybrid SS model (Z-2023) are shown by red solid line and red dashed line, respectively.
In the lower panel, the results for NSs in NS model (AP4) and hybrid NS model (X-2024) are shown by blue solid line and blue dashed line, respectively.
In both panels, $\Delta M/M_\odot=0.04, 0.06, 0.1, 0.2$, increasing downwards.
We choose the masses of pulsars whose $P<15$ ms from~\citet{Ozel2016} and plot their $M$ (with 1-$\sigma$ uncertainties) and $P$ in this $MP$ diagram.

The different effects on parameters of recycled MSPs between SSs and NSs can be seen from Fig.\ref{fig:MP}.
The discrepancy between SSs in LX3630 (red solid line) and NSs in AP4 (blue solid line) is bigger than that between SSs in Z-2023 (red dashed line) and NSs in X-2024 (blue dashed line).
In the low-mass ($M$ below about $1.2 M_\odot$) case, an SS can spin faster than an NS of the same mass by accreting the same amount of mass $\Delta M$, and the difference is more significant when $\Delta M<0.1 M_\odot$.

The constraints on EoS models could be better for masses below about $1.2 M_\odot$ than that for higher masses.
If we can find some fast spinning pulsars with low masses and additionally know the upper limit of $\Delta M$, it would be hopeful to see that whether these pulsars is more likely to be an NS or an SS.
For example, if we can constrain that $\Delta M < 0.04 M_\odot$, then an NS with $M<1.2 M_\odot$ could not be spun up to $P<11$ ms, but an SS could.
Furthermore, we can see that if we only consider SSs in LX3630 model, the discrepancy between SSs and NSs would be more significant.
Certainly, finding more MSPs of $P<15$ ms with low mass and more accurate estimation of $\Delta M$ in the future could help to put more strict constraints on the EoS of pulsars.

\section{Conclusions and discussions} \label{sec:summary}

The global properties of pulsars determined by the EoS, such as mass, radius and the moment of inertia, could play a role in the spin-up process of accreting pulsars in binaries.
We investigate the spin-up of MSPs of NSs (in AP4 and X-2024 models) and SSs (in LX3630 and Z-2023 models).
Combining the spin-up condition and the transferred angular momentum, we can constrain the spin-period and mass of an MSP.
When considering the accretion process, we take into account the evolution of magnetic field during accretion and the fastness-dependent torque.
Our results show that, for the spin-up process of MSPs there could be observable differences between SSs and NSs.

The impeding effect of magnetic field on the spin-up of MSPs would be more significant for NSs than for SSs, especially for the ones with masses below about $1.4M_\odot$.
By accreting the same amount of mass, an SS with mass below about $1.2M_\odot$ can spin faster than an NS of the same mass, implying that it could be easier for an SS to form a low-mass and fully recycled MSP than that for an NS.
Therefore, our results show that the constraints on EoS models would be better in such low-mass case.
For the fast spinning pulsars with low masses, if we can additionally constrain the amount of accreted mass $\Delta M$ (which could be inferred by the nature of donor stars), it would be hopeful to see that whether these pulsars is more likely to be an NS or an SS.

Although there are a lot of work about the recycling of MSPs which mainly focus on the effects of the interactions between pulsars and donor stars during binary evolution, the mass and radius are usually fixed to be typical values, such as $1.4M_\odot$ and 10 km.
We show in this paper, however, that different global structures could bring observational differences, implying that the effects brought by EoS models should be taken into account in studying the recycling of MSPs.

To focus on the EoS-dependent results, it is inevitable to ignore the dependence on other factors in this paper, such as $\phi$ determined by the pattern of disk-fed accretion flow and the critical fastness $\omega_{\rm c}$ determining the accretion torque.
Different choices of $\phi$ and $\omega_{\rm c}$ will influence the quantitative results, shifting the curves in Figs.\ref{fig:spin-up}, \ref{fig:accretion}, \ref{fig:allowed}, and \ref{fig:MP} upwards or downwards, but the qualitative conclusion that an SS could be more easily to form a low-mass and fully recycled MSP than an NS will not change.
In addition, the expressions of the fastness-dependent torque and the evolution of magnetic field during accretion are also model-dependent.
Still, our results indicate that it is worth adding the role of EoS to other effects to study the recycling of MSPs more comprehensively.

The observable differences are manifested in the low-mass case, which can be inferred from the fact that the discrepancy of $M$-$R$ curves between the two SS models and the two NS models is more significant in such low-mass case.
Following the same reasoning, the difference between SSs in LX3630 and NSs in AP4 is more pronounced than that between SSs in Z-2023 and NSs in X-2024.
Although the differences between the two SS models and the two NS models are obvious enough, if we take LX3630 model and AP4 model to represent EoS models of gravity-bound and self-bound respectively, we could get more robust conclusions.

A gravity-bound pulsar, e.g. shown by AP4 EoS, would have nearly unchanged radius when the mass increases from 1 to 2 $M_\odot$, whereas the mass of a self-bound pulsar, e.g. shown by LX3630 EoS, is nearly proportional to the third power of radius when the mass is well below the maximum value.
This difference leads to different evolutions of $r_{\rm m}$ during mass accretion, resulting in different allowed regions in $MP$ diagram of recycled pulsars.
We could make a general conclusion that, it is easier for a self-bound star to form a low-mass and fully recycled MSP than that for a gravity bound star.
Although the we could not yet rule out neither SS nor NS models by the present data, finding more samples of fully recycled low-mass MSPs with accurate mass-measurement and better constraints on the amount of accreted mass in the future, e.g. by SKA and FAST, could be helpful to put more strict constraints on the nature of pulsars.

\section*{Acknowledgements}

We would like to thank Prof. Renxin Xu in PKU, Prof. Wencong Chen in QUT, Prof. Chengjun Xia in YZU and Dr. Chen Zhang in HKUST for useful discussions, and to thank an anonymous referee for useful comments and suggestions. This work is supported by the National SKA Program of China (No. 2020SKA0120300).

\section*{Data Availability}

The data underlying the work in this paper are available upon reasonable request.






\bsp	
\label{lastpage}
\end{document}